# A quantum-dot heat engine operating close to the thermodynamic efficiency limits

Martin Josefsson*, Artis Svilans*, Adam M. Burke, Eric A. Hoffmann, Sofia Fahlvik, Claes Thelander, Martin Leijnse and Heiner Linke+

*NanoLund and Solid State Physics, Lund University, Box 118, 22100 Lund, Sweden*

*These authors contributed equally to this work

+Corresponding author: heiner.linke@ftf.lth.se

**Cyclical heat engines are a paradigm of classical thermodynamics, but are impractical for miniaturization because they rely on moving parts. A more recent concept is particle-exchange (PE) heat engines, which uses energy filtering to control a thermally driven particle flow between two heat reservoirs [1,2]. Because they don't require moving parts, and can be realized in solid-state materials, they are suitable for low-power applications and miniaturization. It has been predicted that PE engines can reach the same thermodynamically ideal efficiency limits as those accessible to cyclical engines [3,4,5,6], but this prediction has never been experimentally verified. Here we demonstrate a PE heat engine based on a quantum dot (QD) embedded into a semiconductor nanowire. We directly measure the engine's steady state electric power output and combine it with the calculated electronic heat flow to determine the electronic efficiency $\eta$. We find that at maximum power conditions $\eta$ is in agreement with the Curzon-Ahlborn efficiency [6,7,8,9] and that the overall maximum $\eta$ is in excess of 70% of Carnot efficiency while maintaining a finite power output. Our results demonstrate that thermoelectric power conversion can in principle be achieved close to the thermodynamic limits, with direct relevance for future hot-carrier photovoltaics [10], on-chip coolers or energy harvesters for quantum technologies.**

Traditional, closed-cycle heat engines are based on an enclosed working medium that exchanges heat, but not particles, with hot and cold thermal reservoirs at temperatures $T_H$ and $T_c$, respectively. The engines' thermal efficiency is bounded by the fundamental Carnot limit $\eta_c = (T_H - T_C)/T_H$ [11]. In practice, however, the goal is usually to optimize efficiency at maximum power: that is, to operate near the lower Curzon-Ahlborn efficiency $\eta_{CA} = 1 - \sqrt{T_C/T_H}$ [7]. The best Stirling engines, for example, reach thermal efficiencies slightly above $0.5\eta_c$ [12], comparable to $\eta_{CA}$ (see Supplementary Information, section A).

A drawback of cyclical engines is that they require moving parts, which severely limits low-power applications, for example in sensors or wearables. By contrast, PE heat engines [2] require no moving elements as they operate by exchanging particles (e.g., photons [1] or electrons [4]) between two heat reservoirs. Theory predicts that PE heat engines can be operated near $\eta_c$ provided that, first, the energy at which particles are exchanged between reservoirs is limited to an energy band much narrower than $kT_H$ [3,5,13], and, second, that said energy is adjusted such that the particle transfer produces no entropy [1,2,3,4,8]. These conditions describe an ideal solid-state thermoelectric system [3,5].

One way to achieve the required energy filtering in solid state is to use a QD that is tunnel-coupled to two electron reservoirs [5,14]. Single-electron orbital states that act as energetically sharp transmission channels (resonances) for electrons at energy $\varepsilon_0$ can be used as energy filters. According to theory, by adjusting $\varepsilon_0$, one can operate the system either near $\eta_c$ [4,14] or near $\eta_{CA}$ [15]. Experimentally, there has been significant progress in the study of the thermoelectric properties of QDs [16,17], and QD-based solid-state cooling devices [18]. However, predictions about the achievable efficiencies in PE heat engines have never been experimentally investigated before as it is challenging to fulfil all requirements for quantitative tests simultaneously: accurate reservoir thermometry; tunable and electrically non-invasive reservoir heating; and a QD that approximates an ideal energy filter.

In this work we explore whether it is possible to reach $\eta_c$ and $\eta_{CA}$ in PE heat engines based on QDs formed by thin InP segments embedded into InAs nanowires (see Fig. 1a), as proposed in [14]. This system offers small and electrostatically stable QDs defined with atomic precision [19] and makes use of well-established device fabrication techniques. The energy-width of the resonance is determined by the tunnel rates $\Gamma$ across the InP segments, and an electrostatic gate can be used to control the resonance energy.

The operation principle of our PE heat engine is illustrated in Fig. 1b. The resonance energy $\varepsilon_0$ is positioned relative to the chemical potentials of the electron reservoirs, $\mu_c$ and $\mu_H$, such that electronic state occupancy at $\varepsilon_0$ is higher in the hot (red) reservoir than in the cold (blue) reservoir. In this configuration, the temperature difference $\Delta T = (T_H - T_c)$ can drive an electric current $I$ against an electrical potential difference $V = (\mu_c - \mu_H)/e$ [4]. In the limit $\hbar\Gamma \to 0$ electrons are transferred only at $\varepsilon_0$. Each transferred electron then produces electric work $eV$ at the cost of removing heat $Q_H = \varepsilon_0 - \mu_H$ from the hot reservoir and depositing $Q_c = \varepsilon_0 - \mu_c$ in the cold reservoir. For finite $\Gamma$, this coupling between charge and heat is no longer exact because of a finite resonance width and effects such as co-tunnelling [20], thus resulting in an increased heat flow. We also note that the existence of a well-defined resonance energy $\varepsilon_0$ requires the single-particle level spacing to be much larger than $kT_H$ to avoid transport through the excited states of the QD.

In the presence of a load $R$ in series with the QD (see Fig. 1c) and zero external bias $V_{ext}$, the circuit self-consistently satisfies the relation $V = -I_{th}R$ where $I_{th} = I(V_{ext} = 0)$ is the thermocurrent. The thermoelectrically produced power in the steady state is $P_{th} = -I_{th}V = I_{th}^2R$ and the efficiency is $\eta = P_{th}/J_Q$, where $J_Q$ is the electronic heat flow leaving the hot reservoir through the QD. We emphasize that $P_{th}$ and $\eta$ depend on $R$, which can thus be used to optimize either $P_{th}$ or $\eta$.

Our experimental device consisted of an InAs/InP nanowire QD in contact with metallic leads as shown in Fig. 1d. We used top heaters [21] for effective thermal biasing of the QD. The differential conductance $dI/dV_{ext} = G$ of the QD as a function of $V_G$ and $V_{ext}$ shows that the QD had a well-defined resonance located at $V_G \approx 0.13$ V, indicated by the intersecting $G$ lines at $V_{ext} = 0$ V (Fig. 2a). This resonance was separated from others by the QD's charging energy of 4.9 meV, which is much larger than $kT_H = 0.17$ meV at the highest electronic temperature $T = 2$ K used in the experiment. No transport via excited states is evident (Fig. 2a). All results discussed in the following were obtained using only this resonance as the energy filter. Data from additional devices are presented in Supplementary Information, section B.

To estimate the engine's efficiency, we calculated the heat flow $J_Q$ based on experimentally determined parameters. This task required a theoretical description that includes full non-linear effects, large electron-electron interactions (Coulomb blockade) and goes beyond the sequential-tunnelling approximation generally used for small $\Gamma$. Our theory approach used the real-time diagrammatic (RTD) technique to expand the Liouville equation for the density matrix in $\Gamma$ and solved the resulting generalized master equations [22,23,24] for a single spin-degenerate energy level. We included all contributions to the current up to order $\Gamma^2$, which includes co-tunnelling, level broadening and energy renormalization processes. We accounted for $R$ by solving the self-consistent equation for $V$ across the QD (Supplementary Information, Section C).

Our analysis of the PE heat engine performance was primarily based on current measurements allowing for accurate determination of $\Gamma$, $T_H$ and $T_C$ (Figs. 2b-d). We determined $\Gamma$ by fitting the RTD theory to the measured peak in $G(V_{ext} = 0)$ as a function of $V_G$ with $\Delta T = 0$ (Fig. 2b). In total four independent measurements were performed at elevated temperatures around 0.5 and 1.0 K to ensure that $kT \gg \hbar\Gamma$, a required condition for the validity of our theory. We assumed equal tunnelling rates across both InP segments which yielded $\Gamma$ values in the range of 8.3 - 9.3 GHz (see Supplementary Information, Sections D and E for details). For further analysis, we used the average value $\Gamma = 8.9$ GHz ($\hbar\Gamma = 5.9$ μeV).

We determined $T_H$ and $T_C$ by measuring $I_{th}(V_G)$ as a response to an applied heater bias $V_{heat}$ (Fig. 2c). The amplitude of $I_{th}$ is sensitive to $T_H$ and $T_C$, and characteristically reverses direction at the resonance ($V_G \approx 0.13$ V) [17]. Using $T_H$ and $T_C$ as free parameters, we found excellent fits of the RTD theory (black lines in Fig. 2c) to the experimental data points, and observed an approximately linear increase of $T_H$ and $T_C$ with $V_{heat}$ (Fig. 2d) (see Supplementary Information,

Section D). We note that the positive and negative $I_{th}$ peak amplitudes are not the same, which is correctly reproduced by our theory. This asymmetry is due to electron-electron interactions in the spin-degenerate QD orbital at $\varepsilon_0$.

Unlike the measured $P_{th}$, the calculated heat flow $J_Q$ does not go to zero at the $I_{th}$ reversal point (see Fig. 3a,b). This is because tunnelling effects of second order in $\Gamma$ effectively decouple the charge current from the heat current, also reducing the maximum achievable efficiency for the PE heat engine. Such effects could be pictured as contra-propagating charges at slightly different energies resulting in $J_Q$ with no net $I_{th}$.

By varying $\varepsilon_0$ we were able to optimize either $P_{th}$ or $\eta$ at each given load $R$ (see Fig. 3c). Maximum $\eta$ was achieved between the peak and the reversal point of $I_{th}$ (black markers in Fig. 3) However, fluctuations in $I_{th}$ lead to significant scatter of the data points in this range and we therefore focused on the maximum $P_{th}$ at each $R$, denoted $P_{max}$, for which the signal-to-noise ratio was better. We found that $P_{max}$ peaked in the $R$ range between 0.7 and 1.5 M$\Omega$, depending on the $V_{heat}$ used, but this value will in general also depend on $\Gamma$ (see Fig. 4a). We note that no simple analytic expression exists for the optimal $R$ for maximum power production [15].

We denote the estimated $\eta$ at $P_{max}$ as $\eta_{Pmax}$. We found $\eta_{Pmax} \approx \eta_{CA}$ for the $R$ that produces the overall maximum $P_{max}$ (Fig. 4b), confirming theoretical predictions [9,13,15]. By instead optimizing $R$ for maximal $\eta_{Pmax}$ yielded efficiencies in excess of $0.7\eta_c$ while still maintaining a finite steady-state power output, roughly equal to one half of the overall maximum power for the same $V_{heat}$ (Fig. 4b). Deviations between the data points and the RTD theory curves for $\eta_{Pmax}$ in Fig. 4b originate from the measured $P_{max}$ being slightly higher than predicted in theory. We note that the sign and magnitude of the underlying deviations in the $I_{th}$ peak values are consistent with a small thermoelectric effect in the contact leads, which our model does not account for.

Our results demonstrate an electronic efficiency at finite power output in excess of 70% of the Carnot limit, comparable to traditional cyclical heat engines (see Supplementary Information, section A), and they confirm that QDs can be operated close to the Curzon-Ahlborn efficiency at maximum power. We achieved this by combining the use of high-quality, epitaxially defined QDs with a novel technique for non-invasive thermal biasing in immediate proximity to the QD [21], and by directly measuring the power produced by the QD as a function of an external load $R$. To determine the electronic heat flow, and thus the efficiency, we used a theory that includes electron-electron interactions (Coulomb blockade), full non-linear effects and higher-order tunnelling. The reliability of the model is validated by its agreement with the experimental data (see Supplemental Information, section D.III). In future work it would be desirable to measure the heat flow directly – a difficult task in a QD heat engine under operating conditions that is subject to both electrical and thermal biases at the same time.

Our experiment approximates "the best thermoelectric" [3] by realizing a system in which particle exchange between heat baths takes place only within a very narrow energy window. Our analysis is limited to electronic thermal reservoirs and does not consider phonon-mediated heat flow, which is a parasitic effect that reduces efficiency [25]. Nevertheless, our results are directly applicable to emerging non-equilibrium devices such as hot-carrier solar cells, which seek to harvest thermal energy stored by photo-generated carriers that are out of thermal equilibrium with phonons [10,26].

**Acknowledgments.** We thank Sebastian Lehmann for structural imaging of the nanowires used in this study. We acknowledge financial support by the People Programme (Marie Curie Actions) of the European Union's Seventh Framework Programme (FP7-People-2013-ITN) under REA grant agreement n°608153 (PhD4Energy), by the Swedish Energy Agency (project P38331-1), by the Swedish Research Council (projects 621-2012-5122, 2014-5490, 2015-00619 and 2016-03824), by the Knut and Alice Wallenberg Foundation (project 2016.0089), Marie Sklodowska Curie Actions, Cofund, Project INCA 600398 and by NanoLund. The computations were performed on resources provided by the Swedish National Infrastructure for Computing (SNIC) at LUNARC.

**Author contributions.** H.L. and M.L. designed and guided the study. E.A.H. and S.F. performed preliminary experiments. S.F. grew the nanowires. A.S., A.M.B. and C.T. designed and fabricated the devices and carried out the experiments. M.J. and M.L. performed the theoretical calculations. M.J. and A.S. analysed the data. All authors contributed to writing and editing the manuscript.

**Competing interests.** The authors declare no competing interests.

**Methods**

*Device specifications*. The QDs in this study are defined in axially heterostructured InAs/InP nanowires [19] (see Fig. 1b) grown by chemical beam epitaxy using gold nanoparticles as catalysts [27,28]. From analyses of transmission electron microscope images of 11 nanowires from the same growth we found that, on average, the diameter of the nanowires is 60 nm, the thickness of the thin InP segments defining the tunnel junctions is 4 ± 3 nm and the length of the InAs QD segment is 17 ± 1.5 nm. The outer InAs segments usable for contacting the nanowire were 350 ± 70 nm and 265 ± 60 nm long. The device fabrication procedure was identical to that reported in [21], with the only difference being that Ti (instead of Ni) was used for electrode adhesion layers. A key element in this architecture is the so-called "top heaters" [21], i.e, heaters fabricated directly on top of the contact leads rather than next to them or using the contact leads themselves as heaters [17]. This geometry allowed us to combine two important features of nanoscale-heaters – higher thermal bias $\Delta T$ with little overall heating of the device and the cryostat, and electrically noninvasive thermal biasing allowing for easy tuning of $\Delta T$. Because $I_{th}$ roughly scales with $\Delta T$, a large $\Delta T$ improves the signal-to-noise ratio of $I_{th}$.

*Measurements*. The experiment was carried out in a Triton 200 dilution refrigerator with RC cold-filtering at 3.5 K with a cut-off frequency of 300 Hz. All measurements were performed with DC using Yokogawa 7651 DC voltage sources for electrical biasing of the QD contacts, the top heater and the global back gate. A Femto DLPCA-200 low-noise current preamplifier with an input impedance of 10 kΩ (at the gain mode of 1 nA/V) was used for measuring the current through the QD. Due to changes in the virtual ground potential of the DLPCA-200 (of the order of 10 µV) whenever the measurement circuit was reconfigured (e.g. when changing $R$ and/or grounding and un-grounding the sample) a more accurate value of $V_{ext}$ zero bias point was determined using $I(V_G)$ of the of the QD's conductance peak as a probe. We used a Femto DLPVA voltage preamplifier to record voltage $V$ (voltage across the QD with the RC filter in series) simultaneously with all $I_{th}$ measurements. This allowed us to characterize drifts in the applied $V_{ext}$ that were smaller than ±1 µV over a period of a single $I_{th}$ measurement trace. Before and after $I_{th}$ measurements at each $V_{heat}$ (at every $R$) four additional measurements of $V(V_G)$ and $I(V_G)$ with $V_{heat} = 0$ V were performed. This allowed us to ensure that the overall $V_{ext}$ drift was less than ±2 µV within the measurements sessions at each $R$. The $V(V_G)$ data (recorded simultaneously with all $I_{th}(V_G)$ measurements) also gave us a consistent set of redundant data with higher noise levels. Due to the filter resistance (4.5 kΩ) of the refrigerator measurement lines, electrical biasing of the top heater was done by setting the two potentials $V^L_{heat}$ and $V^R_{heat}$ on both ends such that the potential of the heater at the device level remains closer to the ground potential to avoid unwanted gating of the QD (see Supplementary Information, beginning of section D for more details). The heater current is approximately $(V^L_{heat} - V^R_{heat})/4.5$ kΩ.

**Data availability.** The data that support the plots within this paper and other findings of this study are available from the corresponding author upon reasonable request.

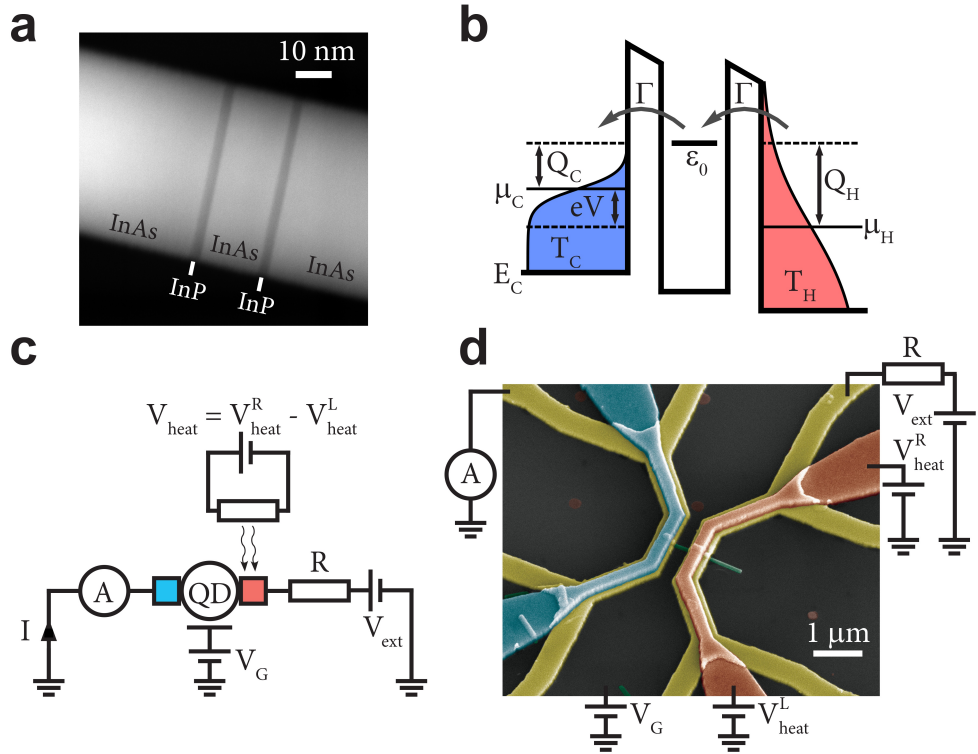

Figure 1: **Experimental device and its operational principle. a** Scanning transmission electron microscopic with high angle annular dark field (STEM-HAADF) image of a representative InAs/InP/InAs/InP/InAs heterostructured nanowire from the same growth as the nanowires in devices. For details on the nanowire dimensions, see Methods. **b** Illustration of a QD-based PE heat engine with resonance energy $\varepsilon_0$. The QD is tunnel-coupled (rate $\Gamma$) to hot and cold electron reservoirs with Fermi distributions characterized by $T_H$, $\mu_H$ and $T_C$, $\mu_C$, respectively. An electron traversing the QD at energy $\varepsilon_0$ removes heat $Q_H$ from the hot reservoir, converts part of it into useful work, $eV$, and deposits the remaining part as heat, $Q_C$, in the cold reservoir. **c** The circuit used for thermoelectric characterization features a tunable resistor $R$ (this also includes 10 k$\Omega$ input impedance of the current preamplifier and 4.5 k$\Omega$ resistance of the RC-filters, not shown), a current preamplifier and a voltage source $V_{ext}$. A separate voltage source $V_{heat}$ = $V^L_{heat} - V^R_{heat}$ is applied in push-pull configuration for running current through a heater that is electrically de-coupled from the hot electron reservoir. **d** False-coloured SEM image of a nominally identical device to the one the experiment is done on. Metallic leads (yellow) make contact to the nanowire (green). Heaters (blue and red) run over the contact leads and are insulated from them by a layer of high-k oxide. One of the heaters (red) is used in the experiment for thermal biasing, while the other (blue) is unused. The resulting $\Delta T = T_H - T_C$ is set by the temperature profile of the phonon bath.

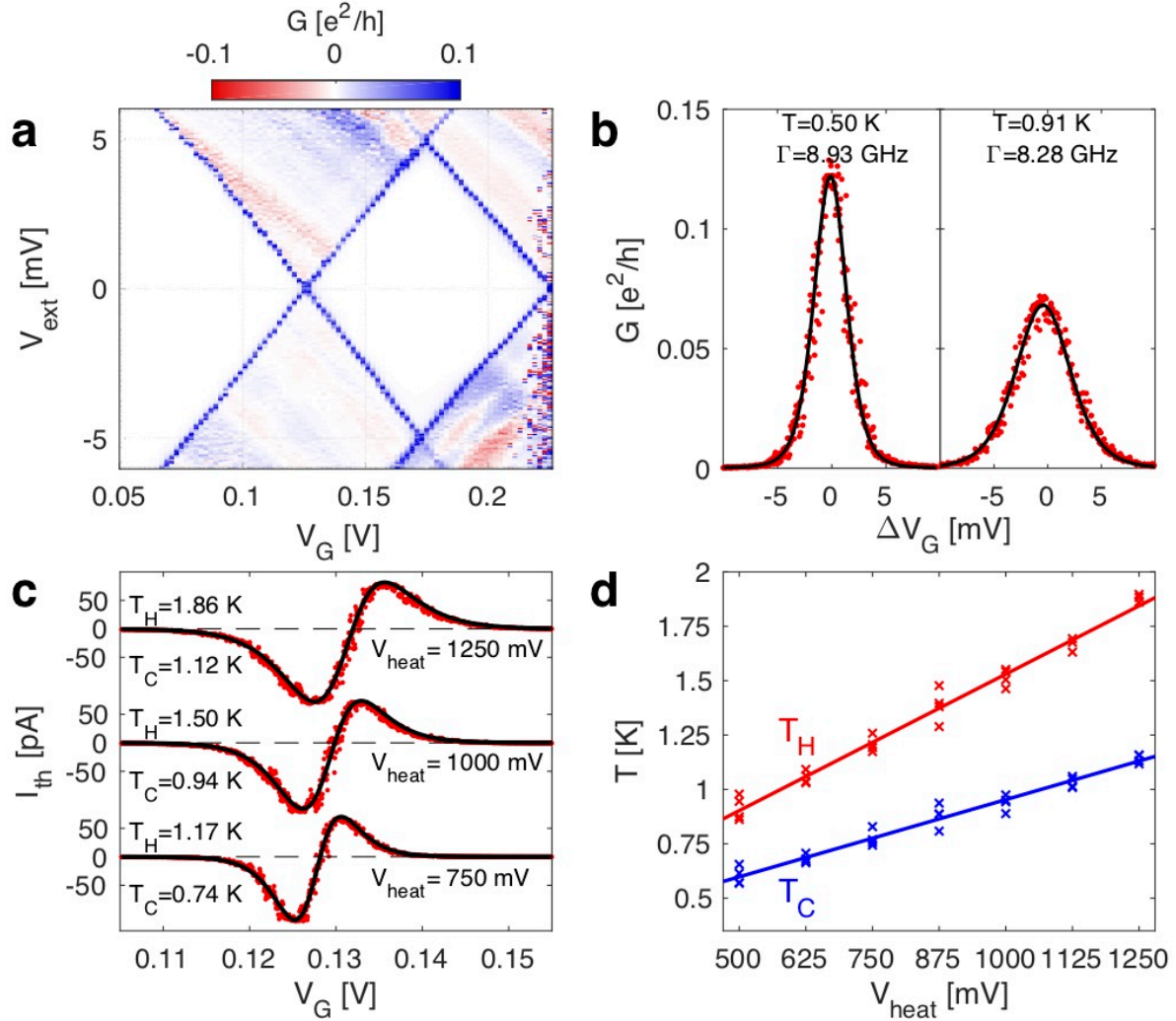

Figure 2: **Electric and thermoelectric characterization of the device.** a Colour plot of differential conductance $G$ of the QD measured at $T < 200$ mK and $V_{heat} = 0$ as a function of $V_{ext}$ and $V_G$. b $G(V_{ext} = 0)$ as a function of $\Delta V_G$ (defined relative to the resonance at $V_G \approx 0.13$ V) at two elevated temperatures close to 0.5 and 1.0 K. Solid lines are fits of the RTD theory with free parameters $T = T_C = T_H$ and $\Gamma$, yielding fit values as reported in the figure. c Measured $I_{th}$ as a function of $V_G$ for different values of $V_{heat}$. Solid curves are fits of the RTD theory where the temperatures $T_C$ and $T_H$ are fitting parameters and $\Gamma = 8.9$ GHz. The shift in the $I_{th}$ reversal point is caused by a parasitic gating effect (see the beginning of section D of the Supplementary Information for more information) d Results from all temperature fits. $I_{th}(V_G)$ is measured four times per each $V_{heat}$, and $T_H$ and $T_C$ are fitted for each individual measurement. All data shown in this figure are taken with $R = 14.5$ kΩ (accounted for in our theoretical approach)

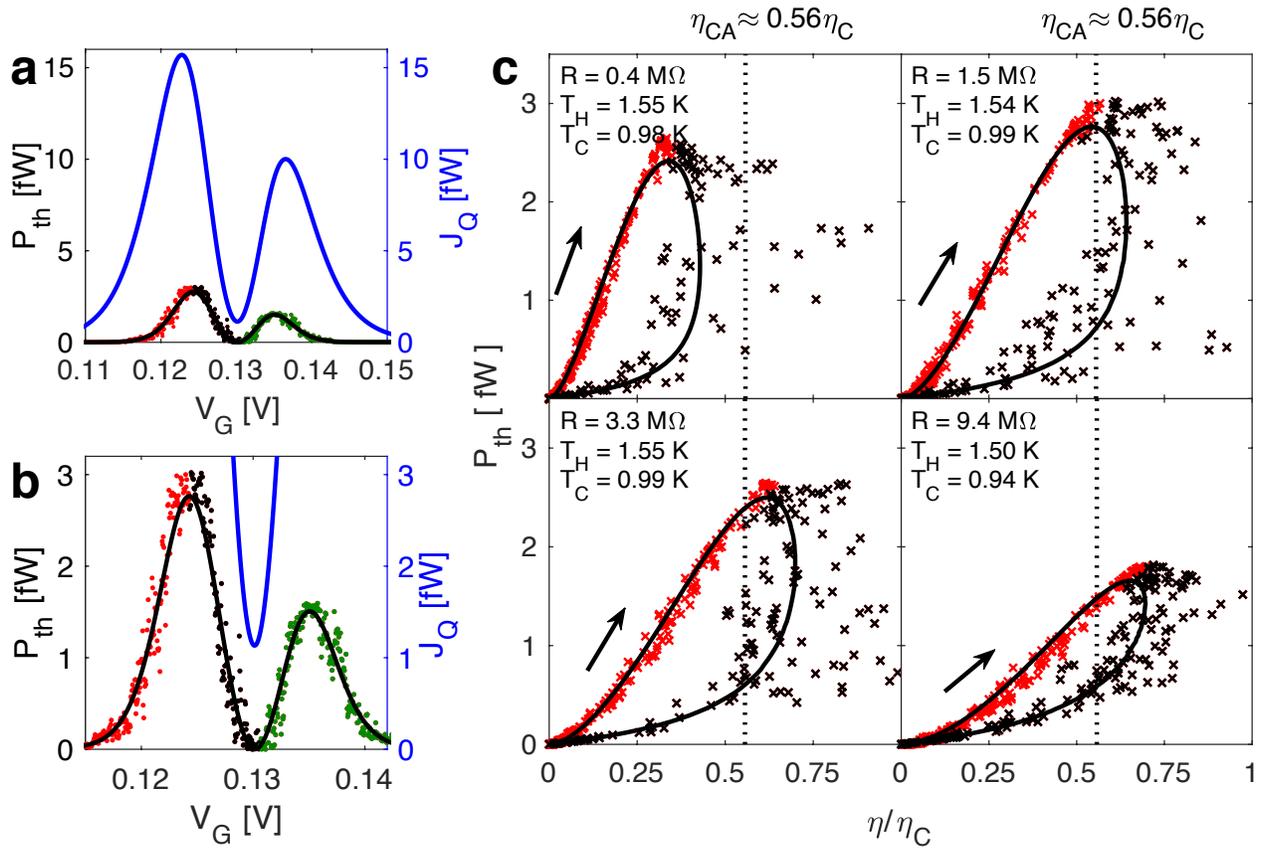

Figure 3: **Thermoelectric performance of the particle-exchange engine. a, b.** Data points are the measured $P_{th}$ at $V_{heat}$ = 1000 mV and $R$ = 1.5 MΩ. The solid lines are the RTD calculations of $P_{th}$ (black) and $J_Q$ (blue) using $T_C$ = 0.99 K, $T_H$ = 1.54 K and $\Gamma$ = 8.9 GHz as obtained from the measurements of $I_{th}$ and $G(V_{ext} = 0$ V) (see Supplementary Information Section D for more details). Data points are coloured to identify different $V_G$ ranges. **c** Parametric plot of $P_{th}$ and $\eta = P_{th}/J_Q$ of the data in **a, b** with $R$ = 1.5 MΩ as well as for three other loads $R$ = 0.4, 3.3 and 9.4 MΩ as indicated. The same colour code as in **a, b** is used (note that the data from the smaller of the two peaks in $P_{th}$, marked with green dots in **a,b,** is not plotted). Red and black markers identify data points from the corresponding $V_G$ ranges as indicated in **a, b.** Data points are based on the measured values of $P_{th}$ and the calculated $J_Q$ using the experimentally determined parameters. The solid lines are based purely on RTD calculations using the same parameters. The arrow indicates the direction for increasing $V_G$. The dashed line is the theoretical prediction for $\eta_{CA} = 0.56\eta_C$ which aligns well with the maximum power $P_{th}$ observed at $R$ = 1.5 MΩ. The spread in data points that are marked in black is due to the fluctuations in the measured current when approaching the $I_{th}$ reversal point in $V_G$. Note that $\eta_{CA}$ is expected to be an approximate upper limit for $\eta$ at the overall maximal power of the engine, which at $V_{heat}$ = 1000 mV used in this figure is $R \approx$ 1.5 MΩ (upper right panel in **c**). For other $R$, where lower powers are achieved, the maximum $\eta$ can be smaller (to left panel in **c**) or larger (lower panels in **c**) than $\eta_{CA}$.

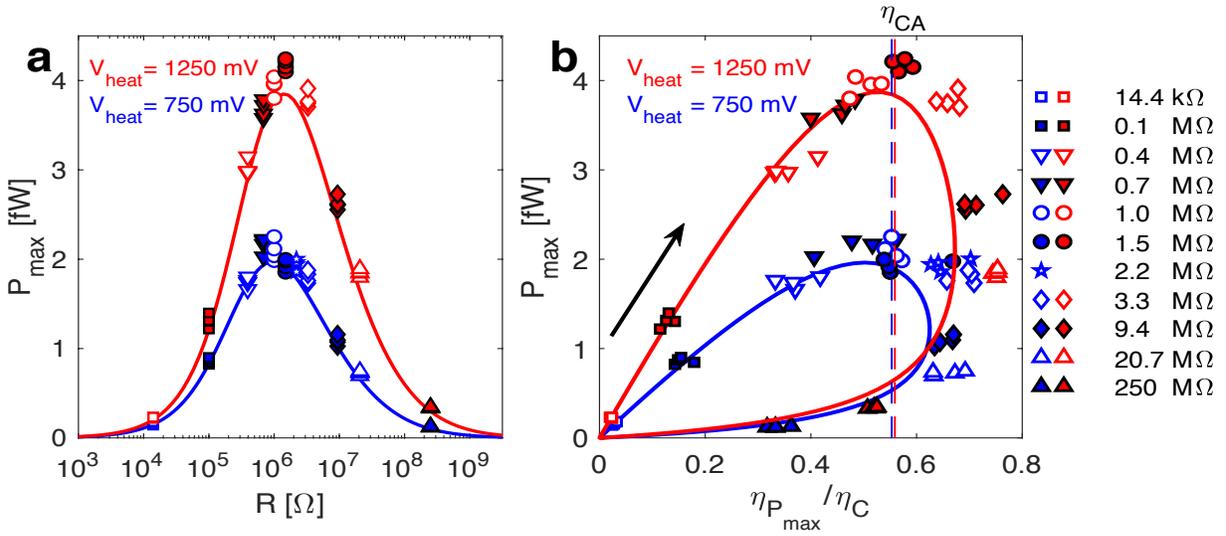

Figure 4: **Device operation at maximum power. a** Power, maximized with respect to $V_G$, $P_{max}$ (markers), compared with the theoretical predictions (solid lines) as a function of the $R$ for two values of $V_{heat}$ as indicated in the figure. Theoretical parameters are $\Gamma = 8.9$ GHz and $T_C = 0.69$ K, $T_H = 1.02$ K for $V_{heat} = 750$ mV and $T_C = 1.13$ K, $T_H = 1.83$ K for $V_{heat} = 1250$ mV as obtained from the fits of the RTD theory to $I_{th}$ and $G(V_{ext} = 0$ V$)$. **b** The same data points for the measured power, $P_{max}$, as in **a,** plotted parametrically for the same range of $R$ values, as a function of the associated $\eta_{P_{max}}$. The solid lines are the theoretical model's predictions. The arrow indicates the direction of increasing $R$ and the vertical dashed lines indicates the corresponding $\eta_{CA}$, both at $\eta_{CA} = 0.56\eta_C$.

# A quantum-dot heat engine operating close to the thermodynamic efficiency limits

# Supplementary information

Martin Josefsson*, Artis Svilans*, Adam M. Burke, Eric A. Hoffmann, Sofia Fahlvik, Claes Thelander, Martin Leijnse and Heiner Linke[+]

## A. Heat engines at various scales

The key results from our work are that (i) we achieve a thermal-to-electric energy conversion efficiency (for electrons) in excess of 70% of the Carnot efficiency, and (ii) an efficiency at maximum power of about the Curzon-Ahlborn efficiency. These are the first tests of the performance limit of particle-exchange heat engines. To place our results into context, in this section we offer an overview of the efficiencies achieved in other systems.

*Power plants.* As conveniently summarized in a table by M. Esposito *et al.* [1] efficiencies for large scale power plants built already decades ago achieved overall efficiencies in the 40% range and correspond to up to 70% of Carnot efficiency. Currently the most efficient heat engine technology is realized in so called combined cycle power plants where the waste heat, produced as a by-product of the gas turbine burning gas at high temperature, is captured and used for boiling water and running another steam turbine. According to a recent review [2] such plants can reach up to 61% overall efficiency and with a simple estimated gas entry temperature (also from Ref. 2) of 1600 °C and ambient temperature of 20 °C, this gives a Carnot efficiency of 84.4 % and the plant efficiency at 72.3% of the Carnot efficiency.

*Stirling engines.* An example of recent developments in Stirling engine technology is NASA's Advanced Stirling Radioisotope Generator program where Advanced Stirling Convertors are developed for integration with General Purpose Heat Sources for generation of electric power [3]. The report (Ref. 3) refers to earlier achievements reported in 2007 when 88 W of AC electric power was achieved with a total efficiency of 38% using 850 °C heater and 90 °C cooler temperatures, which puts it at 56.1% of the Carnot efficiency.

*Thermoelectric materials.* The highest figure of merit *ZT* to date reported for bulk thermoelectric materials (SnSe crystals at 650 °C) is 2.6 [4]. Assuming *ZT* remains at 2.6 when operating between baths at temperatures 650 °C and 23 °C would theoretically allow achieving an overall efficiency of 27.3%, that is 40.4% of Carnot efficiency. Whereas this is significantly lower than the efficiency of existing combined cycle power plants [2] and Stirling engines [3], thermoelectric materials are more suitable for generating small amounts of power [5,6], for example for wearables [7] and self-powered sensors in general [8].

*Quantum dot based cooling devices.* The possibility to use energy-selective electron transport through QDs for cryogenic cooling was first proposed in Ref. 9, recognizing the potential for high efficiency. Experimentally, cryogenic cooling of an electron gas has been realized using the peaked superconducting density of states in superconducting-insulator-normal metal tunnel-junctions [10], using charging effects [11], using QDs [12], and using a single electron transistor (similar to a QD) [13]. However, the electronic efficiency of such devices has to date not been tested.

## B. Reproducibility – additional devices

The data presented in the main text are all based on a single QD device, using a single resonance. This device was chosen because it had very good quality, a value of $\Gamma$ that allowed reaching high efficiency and because we managed to collect the most comprehensive data, including a large range of load resistances that included loads optimized for maximum power production and for maximum efficiency.

The experiments have however been performed on several devices, often using more than one resonance per device. Data on other devices include QDs with stronger as well as weaker tunnel couplings compared with the one in the main text. Fits of the RTD theory to the measured data in order to obtain $\Gamma$ and $T_{C,H}$ were performed for each one of them. The data also includes examples of two energetically close resonances characterized together. Below follows a summary of the other devices used in similar experiments. For every device, the same assumptions regarding $\Gamma$ were made as described in the main text (see Supplemental section D). All nanowires used in fabrication of devices described in this material come from the same growth.

The results from all devices are consistent with the findings presented in the main text. In particular, we show parametric loop graphs for two resonances from device I, which show results consistent with those presented in Fig. 4 of the main text, albeit not for the full range of $R$.

### I. Device I

Two separate resonances were characterized for Device I. Measurements of $I_{th}$ on Resonance 1 were done using loads values $R$ = 1, 2, 3, 3.4, 4, 5, 5.4, 6, 7, 7.4, 8 and 9.4 MΩ. As the $\Gamma$ value for Resonance 2 was found to be much smaller than for any other resonance we measured, the measurements of $I_{th}$ on the Resonance 2 were done using higher loads $R$ = 19.5, 34.4, 44.5, 56.7 and 68.9 MΩ. $G$ peaks for both resonances were characterized before and after $I_{th}$ measurements. The total heater circuit resistance was ≈ 0.6 kΩ such that a relatively high $V_{heat}$ was needed. A typical problem when measuring this device, particularly for resonance 2, was a drifting $V_{ext}$ which made it harder to ensure zero electrical bias of the device. Therefore, measurements of $I(V_G)$ with $V_{heat}$ = 0 mV were taken between each $V_{heat}$ setting. The nonzero $V_{ext}$ showed up in the $I(V_G)$ traces as small $G$ peaks, the magnitude of which was used to estimate a more exact $V_{ext}$ value by fitting it to the RTD theory. The existence of an external bias, $V_{ext}$, also means that the dissipated power in the load is not equal to the produced power of the QD heat engine. Thus, we choose to investigate the engine's effective power, $-IV_{QD} = I(RI + V_{ext})$, and maximum effective power, $I_{max}(RI_{max} + V_{ext})$, instead.

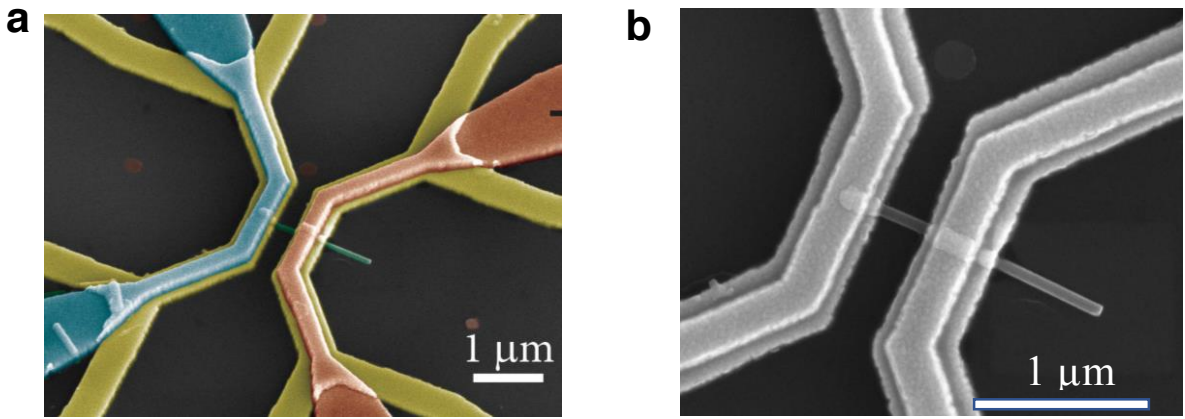

**Supplementary Fig. 1: SEM images of Device I** after the measurements (also used in the main text). **a** Tilted view image with false colours, **b** top view. **Both:** The two parallel metallic (gold) strips, deposited on the substrate oxide, were used as source and drain contact leads. Heater leads are running on top of the source and drain contact leads. Only one of the heaters (coloured in red) was functional at the time of the measurements and was used for thermal biasing, while the other (blue) was not contacted.

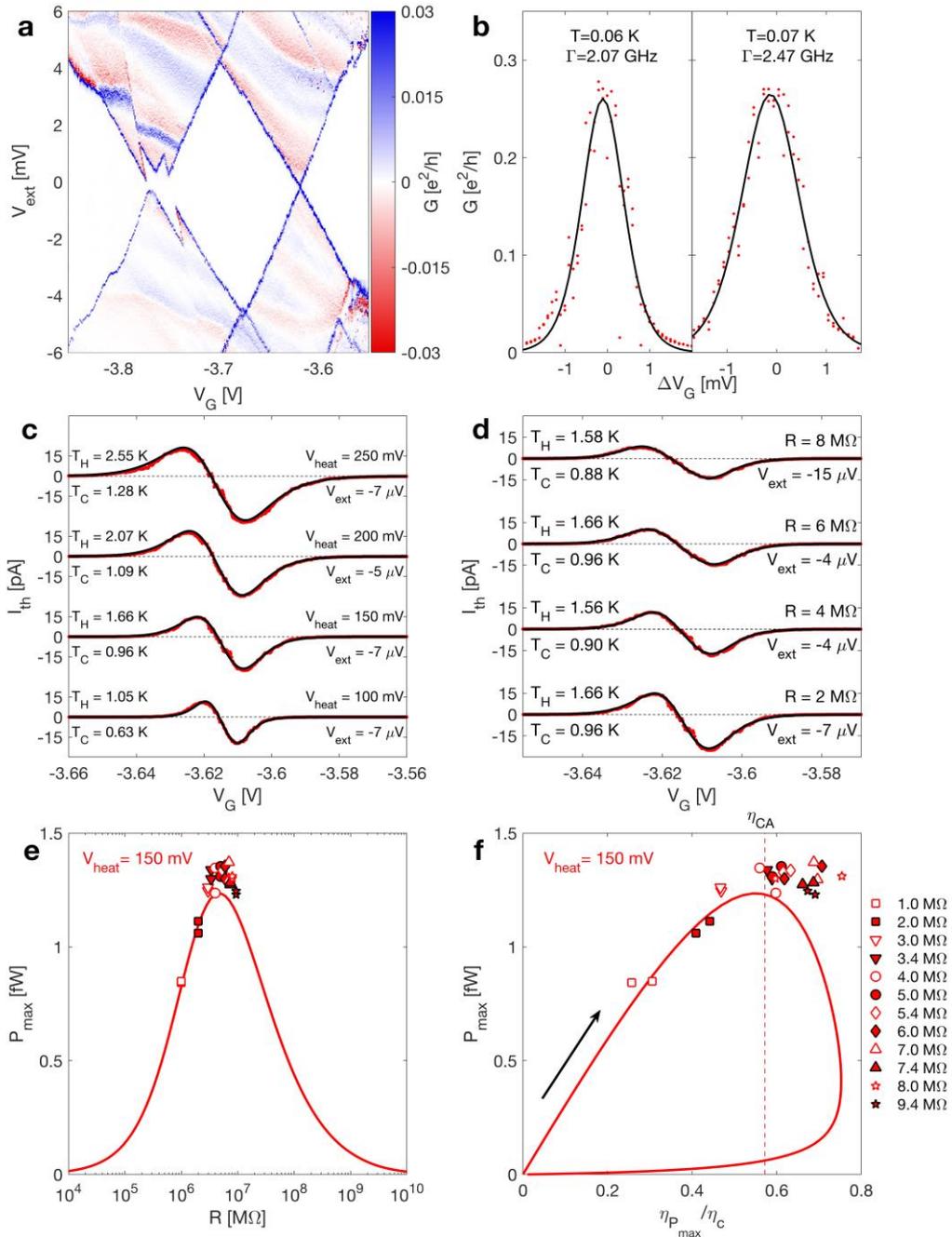

**Supplementary Fig. 2: Characterization of resonance 1. a** Differential conductance $G = dI/dV_{ext}$ as a function of $V_G$ and $V_{ext}$ of Device I, resonance 1. $G$ data is used to determine $E_C = 4.4$ meV and $\alpha_G = 0.032$. **b** The measured $G$ as a function of $\Delta V_G$ (red dots) for the $G$ peak (at $V_G = -3.617$ V), characterized before and after the $I_{th}$ measurements. Fit to the RTD theory of the $G$ peak (black lines) yielding $\Gamma$ values as indicated in the figure. **c** Measured $I_{th}$ as a function of $V_G$ (red dots) for four $V_{heat}$ settings using transport through the same resonance as characterized in b. All four measurements use $R = 2$ MΩ. The corresponding fits of the RTD theory (black curves) use $\Gamma = 2.27$ GHz, $\alpha_G = 0.032$ and small offset $V_{ext}$, as obtained from $G$ peaks at $V_{heat} = 0$ mV (values indicated in the figure). The values for $T_H$ and $T_C$ obtained from the fits are indicated in the figure together with the corresponding $V_{heat}$ settings. **d** Measured $I_{th}$ (red dots) for four different $R$ values (as indicated in the figure) using the same device and $V_{heat} = 150$ mV. The small offset $V_{ext}$, was obtained from $G$ peaks at $V_{heat} = 0$ mV (values indicated in the figure). **e** Measured (markers) and calculated (solid line) maximum effective power, $-V_{QD}I_{max} = I_{max}(RI_{max}+V_{ext})$, plotted against the external load. $V_{ext}$ is obtained from $G$ peaks at $V_{heat} = 0$ mV. **f** Parametric plot of maximum effective power, $-V_{QD}I_{max} = I_{max}(RI_{max}+V_{ext})$, and efficiency at maximum effective power when varying the external load. The arrow indicates the direction of increasing load and the dashed line shows the Curzon-Ahlborn efficiency, $\eta_{CA} \approx 0.57\eta_C$.

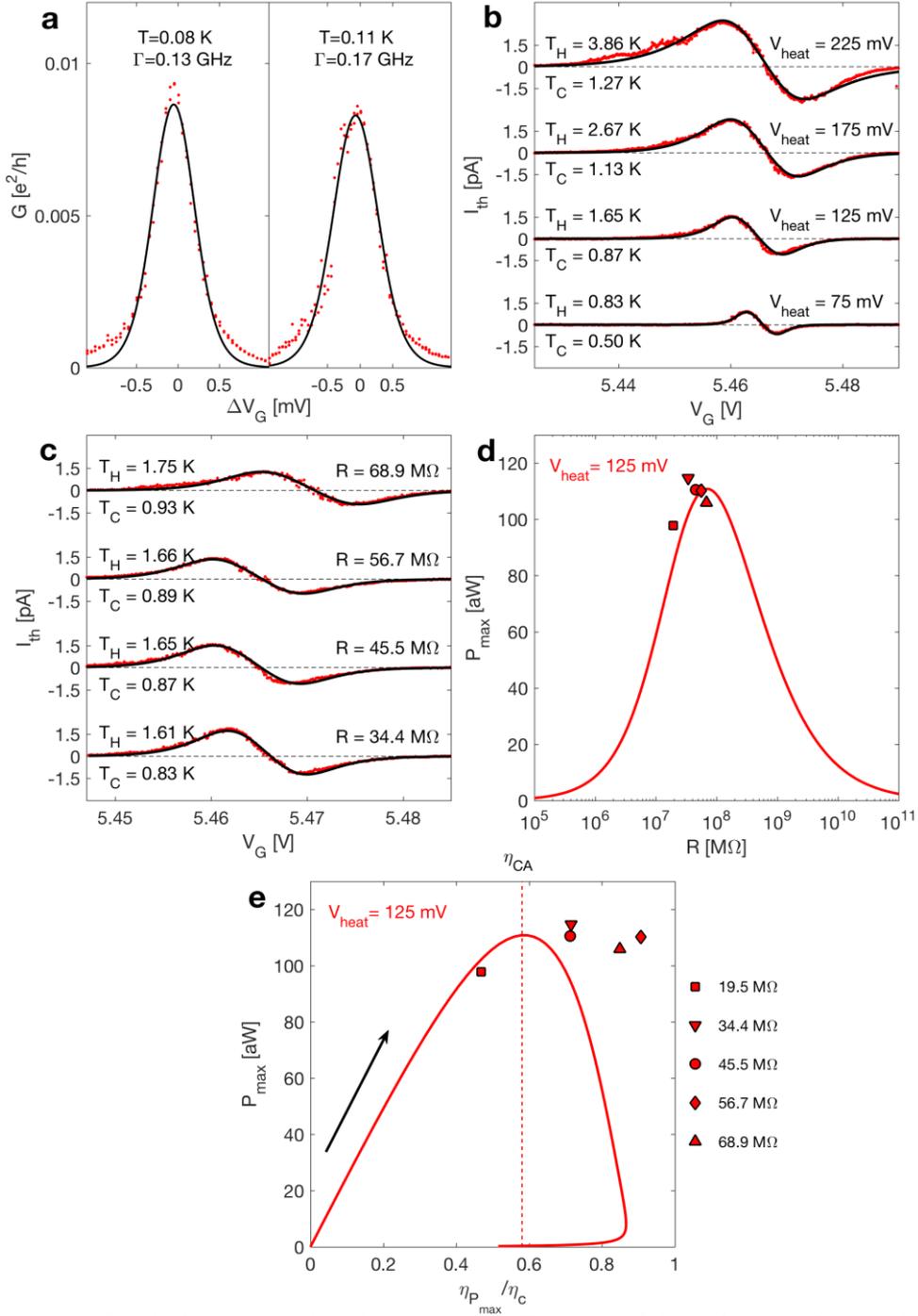

**Supplementary Fig. 3: Characterization of resonance 2. a** The measured $G$ as a function of $\Delta V_G$ (red dots) for the $G$ peak (at $V_G = -5.462$ V) of Device I, resonance 2. This characterization was done after the $I_{th}$ measurements. $\alpha_G = 0.05$ is obtained from characterization of Coulomb blockade peaks at finite $V_{ext}$. Fit of the RTD theory to the measured $G$ peak (black lines) yielding $\Gamma$ values 165 and 170 MHz (before the $I_{th}$ measurements), 132 and 148 MHz (after the $I_{th}$ measurements) giving an average of 154 MHz. **b** Measured $I_{th}$ as a function of $V_G$ (red dots) for four $V_{heat}$ settings using transport through the same resonance as characterized in **a**. All four measurements use $R = 45.5$ M$\Omega$. The corresponding fits of the RTD theory (black curves) use $\Gamma = 154$ MHz, $\alpha_G = 0.05$. The values for $T_H$ and $T_C$ obtained from the fits are indicated in the figure together with the corresponding $V_{heat}$ settings. **c** Measured $I_{th}$ (red dots) for four different $R$ values (as indicated in the figure) using the same device and $V_{heat} = 125$ mV. **d** Measured (markers) and calculated (solid line) maximum power plotted against external load with $V_{heat}$ as indicated in the figure. **e** Parametric plot showing efficiency at maximum power and maximum power when the external load is varied. The solid line represents the theoretical predictions and markers the experimental values. The arrow indicates direction of increasing load and the dashed line shows the Curzon-Ahlborn efficiency, $\eta_{CA} \approx 0.58\eta_C$.

## II. Device II

A set of two resonances was characterized during the measurements on Device II. Several load values were used ($R$ = 0.01, 0.5, 1, 2, 4, 6, 8, 10 MΩ). It can be seen from Fig. S5b that both resonances were likely featuring slightly different tunnelling rates, which lowers the quality of the RTD theory fits to $I_{th}$, Fig. S5c-d, where the average rate of $\Gamma$ = 19.22 GHz was used. To achieve a comparable thermal bias, a relatively large heater bias $V_{heat}$ was needed because of a relatively high heater circuit total resistance ≈ 1.1 kΩ.

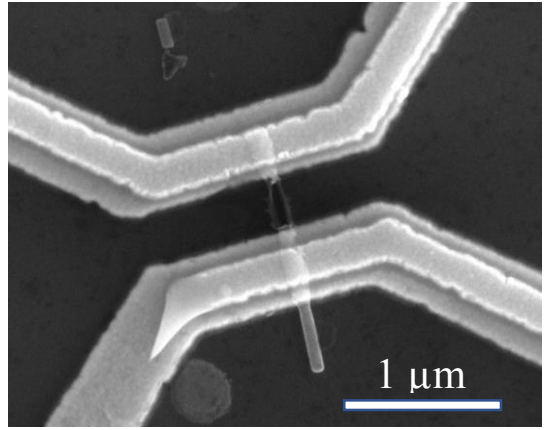

**Supplementary Fig. 4: SEM image of Device II**. The device was damaged in the process of preparing for imaging. The two metallic (gold) strips, deposited on the substrate oxide, were source and drain contact leads. Heater leads are running on top of the source and drain contact leads. Only one of the heaters was functional at the time of the measurements and was used for thermal biasing, the other featured a leak to the back-gate and was kept ungrounded. A shadow-like area was left between the two contact leads where the nanowire containing the QD used to be located.

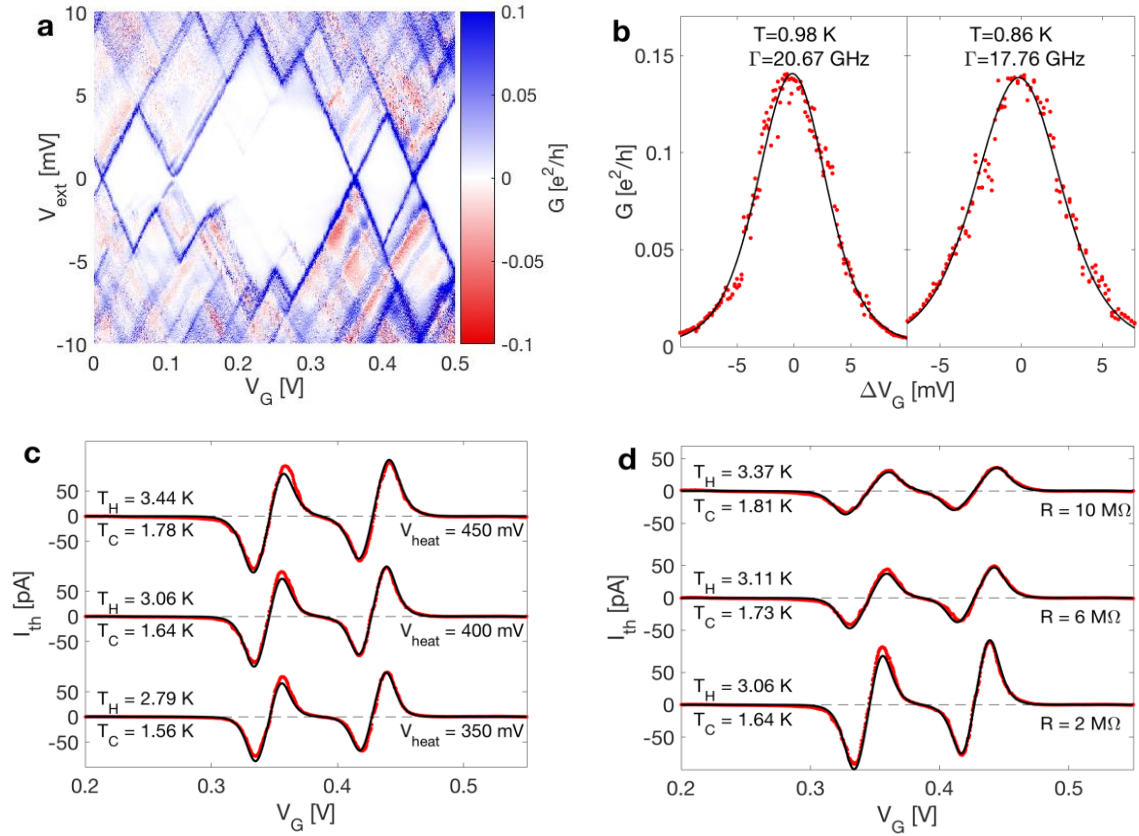

**Supplementary Fig. 5: Characterization of device II. a** Differential conductance $G = dI/dV_{ext}$ as a function of $V_G$ and $V_{ext}$ of Device II. $G$ data is used to determine $E_C = 3.9$ meV and $\alpha_G = 0.048$. **b** The measured $G$ as a function of $\Delta V_G$ (red dots) for two neighbouring $G$ peaks (at $V_G = 0.430$ V and $V_G = 0.347$ V), characterized before the $I_{th}$ measurements. Fits of the RTD theory to the peaks (black lines) yielding $\Gamma$ values, as indicated in the figure. **c** Measured $I_{th}$ as a function of $V_G$ (red dots) for three $V_{heat}$ settings using transport through the same resonances as characterized in **b**. All three measurements use $R = 2$ M$\Omega$. The corresponding fits of the RTD theory (black curves) use $\Gamma = 19.22$ GHz and $\alpha_G = 0.048$. The values for $T_H$ and $T_C$ obtained from the fits are indicated in the figure together with the corresponding $V_{heat}$ settings. **d** Measured $I_{th}$ (red dots) for three different $R$ values (as indicated in the figure) using the same device and $V_{heat}$ 400 mV.

## III. Device III

All measurements on this device were done with $R = 1$ M$\Omega$ which was the input impedance of the current preamplifier SR570. A set of two resonances is characterized twice. Slight charge rearrangements in the device lead to the resonances being shifted in $V_G$ as well as to slight changes in tunnelling rates $\Gamma$. Data from this device is published in Ref.14 without the quantitative analyses of the temperatures and tunnelling rates. The total resistance of the heater circuit was $\approx 115$ $\Omega$.

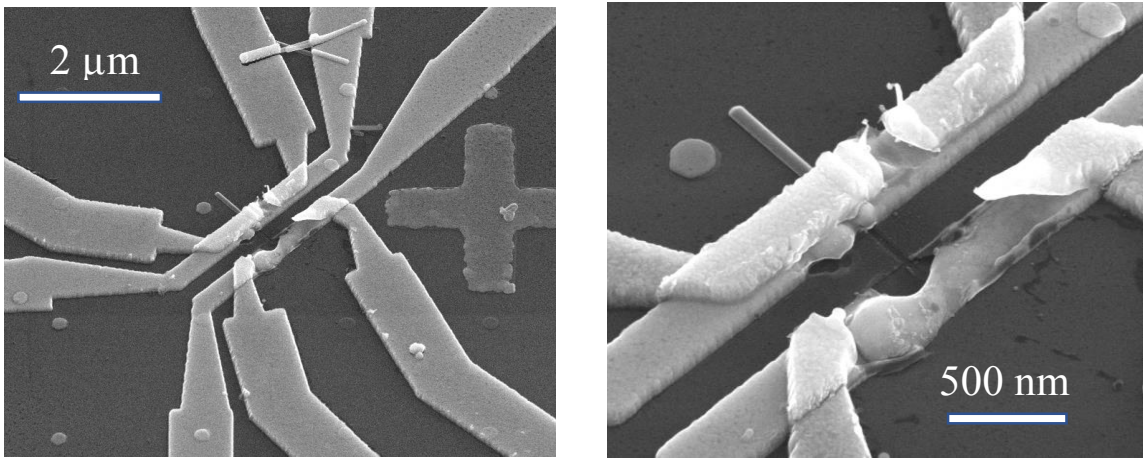

**Supplementary Fig. 6: SEM images of Device III**. Tilted-view SEM images of Device III after the measurements. The device was damaged in the process of preparing for imaging. The two parallel metallic (gold) strips, positioned on the substrate oxide perpendicular to the nanowire, were source train contact leads. Running on top of the source and drain contact leads are the heater leads. Both heaters were functioning at the time of the measurements. Visible in the right-hand picture is a shadow-like area that was left between the two contact leads where the QD-containing nanowire used to be located before it was unintentionally blown away after the experiments, likely by some electrostatic shock.

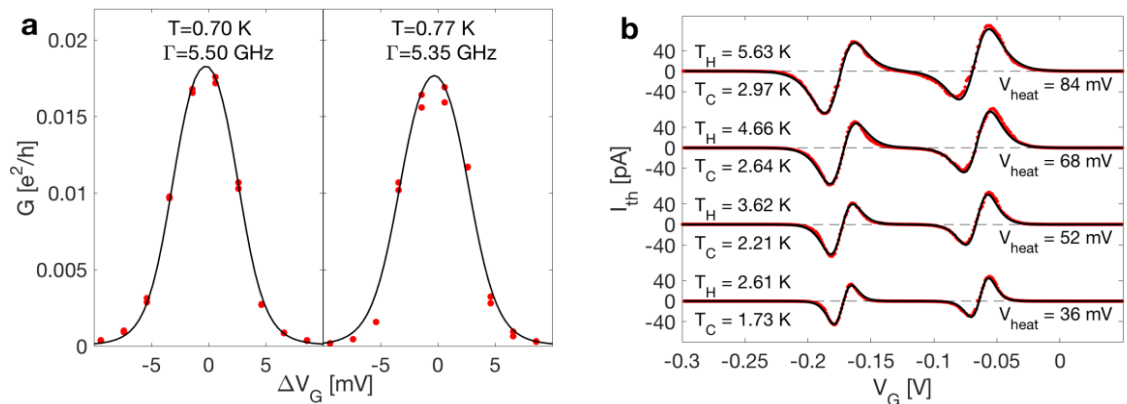

**Supplementary Fig. 7: Characterization 1 of device III**. **a** The measured $G$ as a function of $\Delta V_G$ (red dots) for two neighbouring $G$ peaks (at $V_G = -0.174$ V and $V_G = -0.070$V) of Device III. Fits of the RTD theory to the measured peaks (black lines) yielding similar $\Gamma$ values, as indicated in the figure. $\alpha_G = 0.050$ is obtained from characterization of Coulomb blockade at finite $V_{ext}$. **b** Measured $I_{th}$ as a function of $V_G$ (red dots) for four $V_{heat}$ settings using transport through the same resonances as characterized in **a**. The corresponding fits of the RTD theory (black curves) use $\Gamma = 5.43$ GHz and $\alpha_G$ in a range between 0.050 and 0.052 (set using the $I_{th}$ reversal points in $V_G$). The values for $T_H$ and $T_C$ obtained from the fits are indicated in the figure together with the corresponding $V_{heat}$ settings.

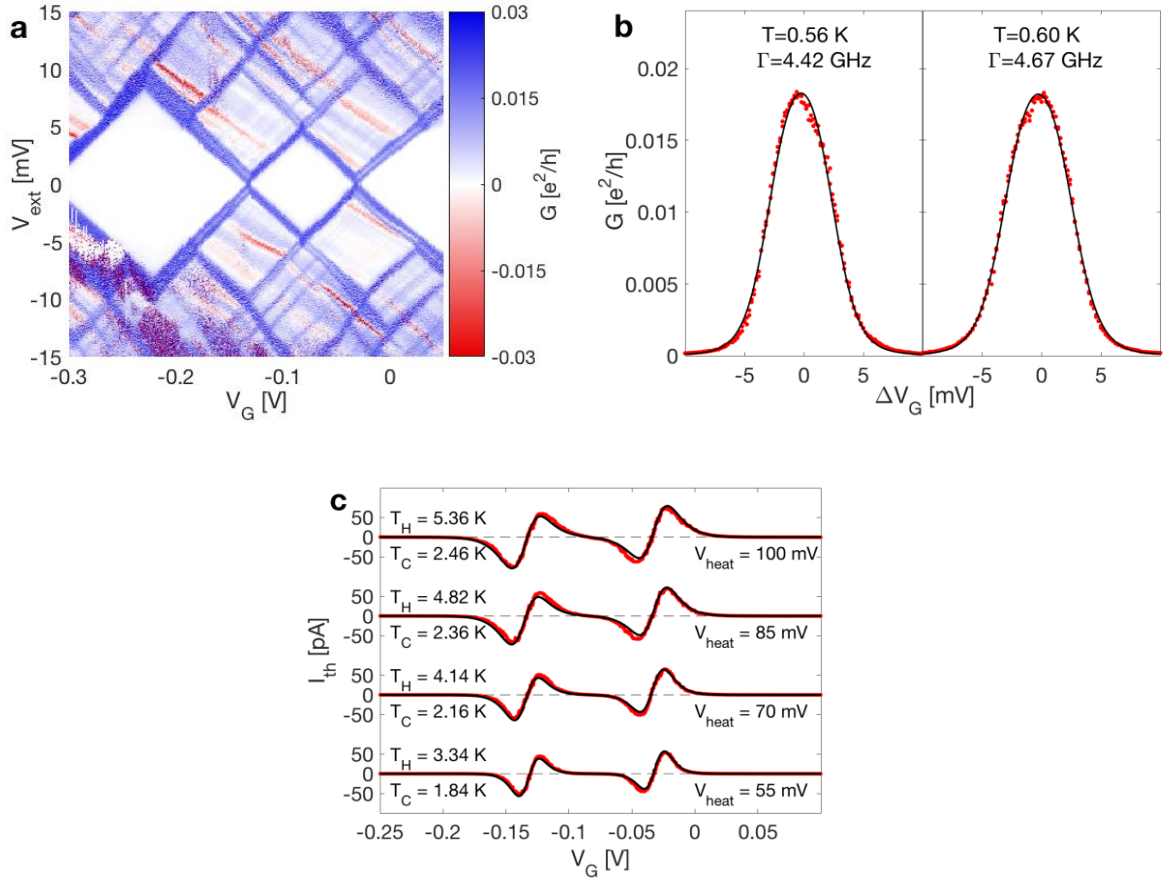

**Supplementary Fig. 8: Characterization 2 of device III**. **A** Differential conductance $G = dI/dV_{ext}$ as a function of $V_G$ and $V_{ext}$ of Device III. $G$ data is used to determine $E_C$ = 4.7 meV and $α_G$ = 0.0472  **b** The measured $G$ as a function of $\Delta V_G$ (red dots) for two neighbouring $G$ peaks at $V_G$ = -0.132 V and $V_G$ = -0.03V, characterized before the $I_{th}$ measurements. Fits of the RTD theory to the peaks (black lines) yielding similar $\Gamma$ values, as indicated in the figure. $G$ characterization before and after $I_{th}$ measurements provided four values $\Gamma$ = 4.67, 4.42, 4.33, 3.40 GHz, yielding an average of 4.21 GHz. **C** Measured $I_{th}$ as a function of $V_G$ (red dots) for four $V_{heat}$ settings using transport through the same resonances as characterized in **b**. The corresponding fits of the RTD theory (black curves) use $\Gamma$ = 4.21 GHz and $α_G$ in a range between 0.047 and 0.048 (set using the $I_{th}$ reversal points in $V_G$ that changed slightly with temperature). The values for $T_H$ and $T_C$ obtained from the fits are indicated in the figure together with the corresponding $V_{heat}$ setting.

## IV. Device IV

Device IV was the device used for the experiments in the main paper. In Fig. S9 we show SEM images of the device taken after the measurements.

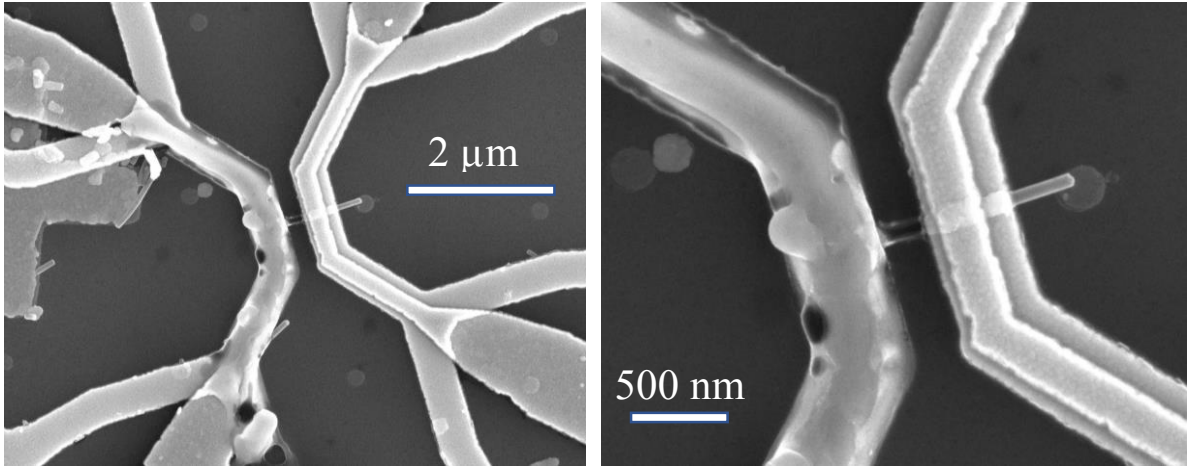

**Supplementary Fig. 9: SEM images of Device IV**. Top view SEM images of Device IV after the measurements. The device was damaged in the process of preparing for imaging. The two parallel metallic (gold) strips, deposited on the substrate oxide perpendicular to the nanowire, were used as source and drain contact leads. The leads running on top of the source and drain contact leads are the heater leads. Only one of the heaters was functional at the time of the measurements and was used for thermal biasing, the other featured a leak to the back-gate and was kept ungrounded. A shadow-like area was left between the two contact leads where the nanowire containing the QD used to be located.

# C. Theory

Below we describe the theoretical method used in the main paper. In short, our theory is based on the real-time diagrammatic techniques originally developed in Refs. [15,16]. To model the experiment we need to consider a finite voltage bias as well as a temperature difference, and we calculate both the charge and heat current, for which most aspects of the underlying theory have been developed in Refs. 17,18,19.

## I. Hamiltonian

We are interested in open quantum systems where a central quantum dot (QD) is coupled to macroscopic reservoirs via hybridization terms in the Hamiltonian. The Fock-space of the central QD is typically small. The reservoirs are assumed to be described by the grand canonical ensemble.

The total system is modelled as the Hamiltonian

$$H = H_D + \sum_r H_r + \sum_r H_{T,r} = H_D + H_R + H_T. \qquad (1)$$

Here, $H_D$ is the Hamiltonian describing the QD, which is assumed to be diagonal in the many body states of the QD

$$H_D = \sum_\sigma \varepsilon_\sigma n_\sigma + E_C n_\uparrow n_\downarrow, \qquad (2)$$

with single particle energy $\varepsilon_\sigma$ for an electron with spin $\sigma =\uparrow,\downarrow$, and energy cost $E_C$ associated with double occupation (because of electron-electron interactions). $n_\sigma = d_\sigma^\dagger d_\sigma$ is the number operator made up from the fermion creation ($d_\sigma^\dagger$) and annihilation ($d_\sigma$) operators. In the following, field operators acting on the QD subspace are denoted by the letter $d$ and those acting on the reservoir subspace by the letter $c$. A reservoir $r$ is described by

$$H_r = \sum_{k,\sigma} \omega_{k,\sigma,r} n_{k,\sigma,r}, \quad n_{k,\sigma,r} = c_{k,\sigma,r}^\dagger c_{k,\sigma,r} \qquad (3)$$

where $\omega_{k,\sigma,r}$ is the eigenenergy for a state in reservoir $r$ with quantum numbers $k, \sigma$. Finally, the QD and reservoir $r$ are coupled by the tunnelling Hamiltonian

$$H_{T,r} = \sum_{k,\sigma} t_{k,\sigma,r} d_\sigma^\dagger c_{k,\sigma,r} + h.c.. \qquad (4)$$

The amplitude for electron tunnelling is given by $|t_{k,\sigma,r}|^2$, which allows us to define a tunneling rate

$$\Gamma_r = \frac{2\pi \nu_r |t_{k,\sigma,r}|^2}{\hbar}. \qquad (5)$$

Here $\nu_r$ is the density of states of the reservoir and $\hbar$ is Planck's reduced constant. In the sections to come we utilize the wide band approximation assuming that $\nu_r$ is constant over an energy $D$ much larger than any other involved energy scale, and for convenience we set $\hbar = e = k_B = 1$.

## II. Liouville-von Neumann equation

Our aim is to calculate the nonequilibrium state of the QD when the reservoirs are kept at different temperatures and chemical potentials, and then to find the charge and energy currents flowing as a result of this nonequilibrium condition. The starting point is the Liouville-von Neumann equation for the time evolution of the full system density matrix ($\rho$)

$$\frac{d}{dt}\rho = -i[H,\rho]_- = -iL\rho, \tag{6}$$

where $L \bullet \equiv [H,\bullet]_-$ is the Liouville super-operator acting on an arbitrary operator $\bullet$ (we similarly define $L_T \bullet \equiv [H_T,\bullet]_-$ and $L_D \bullet \equiv [H_D,\bullet]_-$). Here the term super-operator is used for mathematical objects operating on normal operators. The matrix elements of such a super-operator are evaluated as

$$A_{cd}^{ab} = \langle c|(A|a\rangle\langle b|)|d\rangle. \tag{7}$$

There are many ways to find an approximate solution to eq. 6, we here use the real time diagrammatic technique [15] formulated in Liouville space [20]. The first step is performing a Laplace transformation and expanding the resulting expression in $L_T$. This expansion can then be re-summed by identifying it as a geometrical series to obtain

$$\rho_D(z) = \frac{i}{z - L_D - W(z)} \rho_D(0), \tag{8}$$

where $\rho_D = \text{Tr}_R\{\rho\}$ is the QD (reduced) density matrix and the kernel $W(z)$ is given by

$$W(z) = \sum_{k=1}^{\infty} \text{Tr}_R (L_T \frac{1}{z - L_D - L_R})^k L_T \rho_R|_{irred.}. \tag{9}$$

*irred* here refers to the irreducible diagrams in the perturbative series [15,20,21].

Here, we are only interested in the stationary state (long time limit and time-independent Hamiltonians). Re-arranging eq. 8 and taking the stationary (zero frequency) limit yields

$$0 = (-iL_D + W)\rho_D. \tag{10}$$

The generalized master equation in eq. 10 is valid for the full QD density matrix, but for our model the QD density matrix will be diagonal in spin and charge because these quantities are conserved by the total Hamiltonian.

Next we will introduce a convenient super-operator notation in Liouville space which will make the evaluation of $W$ fairly straight forward.

## III. Kernel evaluation

In order to evaluate the series in eq. 9 it is convenient to find super-operators in Liouville space whose properties resemble those of fermion field operators in Fock space.

We define the notation for our fermionic creation and annihilation operators acting on the QD subspace as

$$d_{\eta\sigma} = \begin{cases} d_\sigma & \text{if } \eta = - \\ d_\sigma^\dagger & \text{if } \eta = + \end{cases}. \tag{11}$$

We combine all indices to a single index represented by a positive integer, $1 = \eta_1\sigma_1$, and denote by a bar the sign change of appropriate numbers i.e. $\bar{1} = \bar{\eta}_1\sigma_1 = (-\eta_1)\sigma_1$ and $\bar{q}_1 = -q_1$. Furthermore, a sum over such a multi-index ($\Sigma_1$) is implicitly understood as sums over all involved indices if nothing else is stated. The field operators acting on the reservoir subspace are defined in a similar manner where the indices $k$ and $r$ are also included in the multi-index.

Using the notation in eq. 11 we define Liouville field super-operators as

$$\mathcal{G}_1^{q_1} \bullet = \frac{1}{\sqrt{2}}(d_1 \bullet + q_1(-1)^N \bullet (-1)^N d_1), \quad q_1 \in \{-,+\} \tag{12}$$

with $N = \sum_\sigma n_\sigma$. These operators have properties similar to those of fermionic operators in Hilbert space

$$(\mathcal{G}_1^{q_1})^\dagger = \mathcal{G}_{\bar{1}}^{\bar{q}_1}, \tag{13}$$

$$[\mathcal{G}_2^{q_2}, \mathcal{G}_1^{q_1}]_+ = \delta_{q_2\bar{q}_1}\delta_{2\bar{1}}\mathcal{J}. \tag{14}$$

Analogously for the field operators acting on the reservoir subspace we use the notation $\mathcal{J}_1^{q_1}$.

Using these operator definitions the kernel in eq. 9 can be evaluated. However, this expression contains an infinite sum of terms and needs to be truncated. Only terms of even order in $k$ gives a non-vanishing contribution and we include terms up to order $k = 4$, which account for processes $\propto \Gamma$ and $\Gamma^2$. The remaining terms are evaluated by collecting all reservoir super-operators and calculating their expectation value using Wick's theorem. The leading order terms of the kernel now reads

$$W^{(2)} = \sum_{q_1 1} \frac{\Gamma_1}{2\pi} \mathcal{G}_{\bar{1}}^+ \frac{q_1\gamma_1^{q_1}}{i0 - L_D + \eta_1\omega_1} \mathcal{G}_1^{\bar{q}_1}, \tag{15}$$

and the next to leading order [17,18,19].

$$W^{(4)} = \sum_{12q_1q_2} \frac{\Gamma_1\Gamma_2}{(2\pi)^2}(\mathcal{G}_{\bar{1}}^+ \frac{1}{\eta_1\omega_1 + i0 - L_D}\mathcal{G}_{\bar{2}}^+ - \mathcal{G}_{\bar{2}}^+ \frac{1}{\eta_2\omega_2 + i0 - L_D}\mathcal{G}_{\bar{1}}^+)$$

$$\times \frac{\bar{q}_2\gamma_2^{\bar{q}_2}}{\eta_1\omega_1 + \eta_2\omega_2 + i0 - L_D}\mathcal{G}_2^{q_2} \frac{\bar{q}_1\gamma_1^{\bar{q}_1}}{\eta_1\omega_1 + i0 - L_D}\mathcal{G}_1^{q_1}, \tag{16}$$

Where

$$\gamma^{q_1} = \delta_{q_1+} + \delta_{q_1-}\tanh(\frac{\eta_1(\omega_1 - \mu_1)}{2T_1}). \tag{17}$$

The resulting integrals in equation 15 and 16 are solved in the Supplementary Information of Ref.17.

## IV. Observables

Any observable can now be calculated using the density matrix

$$\langle A \rangle = \text{Tr} A \rho. \tag{18}$$

When evaluating this expectation value it is however possible to make several simplifications by writing

$$\langle A \rangle = \frac{1}{2} \text{Tr}\, L_A^+ \rho, \quad L_A^+ \bullet = [A, \bullet]_+. \tag{19}$$

### IV.a Charge current

The charge current is given by the time derivative of the particle number in one of the reservoirs

$$I_r = -\frac{d}{dt} N_r = -i[H, N_r], \quad N_r = \sum_{k,\sigma} n_{k,\sigma,r}. \tag{20}$$

The calculation of the charge current can be simplified by using the fact that the charge is conserved in all tunnelling processes [22]. Using this conservation law the current leaving reservoir $r$ is given by

$$I_r = -\frac{1}{2} i\, \text{Tr} L_N^+ W_r \tag{21}$$

where $N = \sum_\sigma n_\sigma$ and $W_r$ is similar to $W$ in eq. 9 with the only difference that the left-most $L_T$ is replaced by $L_{T,r}$.

### IV.b Energy and heat current

A similar treatment as the charge current is possible for the energy current. It is however non-trivial because the tunnelling Hamiltonian introduces additional dynamics which cannot be ignored [19, 23]. The energy current thus consists of two parts

$$J_{E_r} = i \text{Tr}_D L_{H_D}^+ W_r \rho_D - i \text{Tr}_D W_{\Gamma,r} \rho_D \tag{22}$$

where the first term on the right-hand side is evaluated in the same manner as eq. 21. In order $\Gamma$ only the first term in 22 contributes and in the next to-leading order the second term is given by [19]

$$-i \text{Tr}_D W_{\Gamma,r} \rho_D = \frac{\Gamma_r \Gamma_{r'}}{2\pi} \sum_{\sigma, p=\pm} [f_{r\sigma p} + p f_{r\sigma p} \text{Tr}_D (-\mathcal{I})^{n_{\sigma'}} \rho_D], \tag{23}$$

where $r' \neq r$, $\mathcal{I}$ the identity matrix and

$$f_{r\sigma p} = \text{Re}\left[\Psi\left(\frac{1}{2} - i\frac{\frac{E_C}{2(p-1)} - \epsilon + \mu_r}{2\pi T_r}\right) - \Psi\left(\frac{1}{2} - i\frac{\frac{E_C}{2(p-1)} - \epsilon + \mu_{r'}}{2\pi T_{r'}}\right)\right], \quad (24)$$

with $\Psi$ denoting the digamma function.

The heat flow leaving the hot reservoir is related to the above quantities through the first law of thermodynamics

$$J_{Q_H} = J_{E_H} - \frac{\mu_H}{e}I_H. \quad (25)$$

## V. Self-consistent solution

The introduction of a serial load $R$ introduces additional computation steps because the electrical bias $V_{QD}$ across the QD is set by the current flowing through the dot, $V_{QD} = I_{QD}R$, rather than by an external bias applied to the reservoirs (see circuit diagram in Fig. 1c in the main paper). $V_{QD}$ thus needs to be obtained by solving the self-consistent equation

$$I_{QD}(V_{QD}) - \frac{V_{QD}}{R} = 0, \quad (26)$$

which is done numerically in an iterative manner. $I_{QD}(V_{QD})$ is obtained from eq. 21 with $eV_{QD} = \mu_H - \mu_C$. This means that for each iteration in the root finding algorithm (eq. 26) $H_R$ is changed and thus a new density matrix and new currents need to be calculated by going through the steps described above.

# D. Details of the measurement and fitting process

The parameters $\Gamma$, $T_C$ and $T_H$ were obtained from fitting the RTD theory to the measured conductance and current using least square fits. Data in Fig. 2a in the main text allowed determination of the gate coupling $\alpha_G = 0.049$ which was further used as an input parameter. Motivated by the straight Coulomb diamonds (see Fig. 2a in the main paper) we let the potential $eV$ drop symmetrically over both tunnel barriers. Furthermore, the value of $\varepsilon_0$ was adjusted to compensate for electrostatic potential variations, not considered by the model. On a measurement-to-measurement basis, one source of such variations in general is charge rearrangements in the proximity of the QD. In our experimental setup there was an additional factor, namely finite electrical potential changes at the top-heater as a function of $V_{heat}$. The origin of this effect can be understood in the following way. Whenever the heater is biased by $V_{heat}$ in order to run a heating current through it, the filter resistances in series with the heater on both sides develop voltages across them. In case this effect is not counterbalanced by applying $V_{heat} = V^R_{heat} - V^L_{heat}$ in a push-pull manner (by setting positive bias $V^R_{heat}$ on one side of the heater and negative bias $V^L_{heat}$ on the other, or visa versa) it results in an overall potential change at a heater electrode that can slightly gate the QD. A characteristic of this effect is that this parasitic "gating" scales linearly with $V_{heat}$. Unfortunately, due to different resistances of the voltage sources' output filters the balancing of the potentials was not done perfectly and a slight shift in the $I_{th}$ reversal point with increasing $\Delta T$ remained and was compensated for when fitting.

## I. Tunnel rate characterization

The tunnelling rate $\Gamma$ can be related to the $G(V_{ext} = 0)$ of the Coulomb peaks (Fig 2b in the main text). The $V_G$ dependence of the $G$ peaks is sensitive to $\Gamma$ and temperature $T$. Experimentally we determined $G(V_G)$ by measuring $I$-$V_{ext}$ curves and finding linear fits at each $V_G$. The measurements were done in DC and used the range of $V_{ext}$ in which the $I$-$V$ curves are linear, typically $V_{ext} \leq \pm 50$ μV. We believe that most of the scatter of the data points in $G$ originates from the variations in the electrostatic potential landscape due to charge rearrangements in the proximity of the QD, having an effect similar to $V_G$ noise.

The explained $G$ peak characterization was done at the beginning as well as at the end of the thermoelectric characterization of the resonance. For better consistency $G(V_G)$ was measured at the cryostat base temperature (estimated electron temperature slightly below 0.2 K) as well as at two elevated temperatures obtained by rising the cryostat mixing chamber temperature to 0.5 and 1.0 K. From the resulting fits of the RTD theory to the measured data we could conclude that at the cryostat base temperature $\hbar\Gamma/kT \approx 0.34$ was too large for the theory to quantitatively reproduce the measurements with high accuracy because we were not deep enough in the perturbative regime ($\hbar\Gamma/kT \ll 1$). Thus only $G$ measurements at elevated temperatures were used for determining $\Gamma$. The four remaining conductance measurements yielded $\Gamma = 8.93$ GHz (and $T = 503$ mK), $\Gamma = 9.25$ GHz (and $T = 480$ mK), $\Gamma = 8.28$ GHz (and $T = 906$ mK) and $\Gamma = 9.06$ GHz (and $T = 953$ mK). This gives an average $\Gamma = 8.88 \pm 0.42$ GHz which was used for all calculations in the main text. One standard deviation is used as the error interval.

## II. Temperature characterization

In our experimental device the electron reservoirs were, effectively, the short (~ 100 nm) InAs nanowire segments, and our analysis requires determination of their electronic temperatures $T_H$ and $T_C$. Fitting the RTD theory to the characteristic behaviour of $I_{th}(V_G)$ is a non-invasive method with good sensitivity given that the thermal bias $\Delta T = T_H - T_C$ is big enough to produce a sufficiently large $I_{th}$ for accurate measurements. We preferred using $I_{th}(V_G)$ over the more conventionally used $I(V_{ext})$ [24], because of the energy dependency of the tunnelling rates at higher $V_{ext}$ conditions (manifesting in a $V_{ext}$ dependent saturation current Fig. S13). This is likely due to the small size of the nanowire lead segments that affect the density of states in the reservoirs. Using the $I_{th}(V_G)$ ensures that the temperatures are characterized at the conditions in which the engine itself is characterized.

$T_H$ and $T_C$ were determined from least-square fits of the RTD theory to $I_{th}(V_G)$ using the temperatures as the only two free parameters. We used $\Gamma$ = 8.88 GHz and $\alpha_G$ = 0.049 as determined from the $G$ measurements.

Small measurement-to-measurement variations in the $V_G$-location of the resonant energy $\varepsilon_0$ were compensated for by setting the theoretical $I_{th}$ reversal point equal to the measured one before fitting. We also accounted for a small offset in the current (ca -0.3 pA), determined from the current reading in the Coulomb blockade regime without any biases, $V$ or $\Delta T$, applied. The external load was included in the model by self-consistently solving for $V$ across the QD.

$I_{th}(V_G)$ was measured four times for each $V_{heat}$ configuration (two times sweeping $V_G$ in one direction, and two in the other). The two-parameter RTD theory fit for $T_H$ and $T_C$ was done for each $I_{th}(V_G)$ trace independently. An example of four such fits is shown in Fig. S10a. This procedure was repeated for all 11 $R$ values used in the experiment of the presented device and similar temperatures for all $R$ values were found. For an example, the Fig. S10b shows the result for four $I_{th}$ using $V_{heat}$ = 1000 mV, but for different $R$. In Fig. S10c we summarize thermometry results performed at 7 different thermal bias conditions 4 times for each of the 11 loads. Overall both $T_H$ and $T_C$ are found to increase approximately linearly with $V_{heat}$.

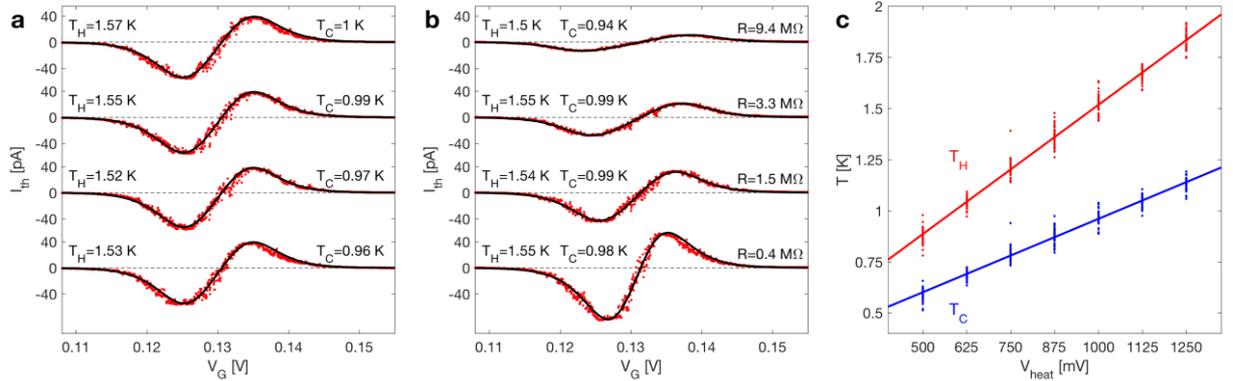

**Supplementary Fig. 10: Temperature characterization.** The fits of the RTD theory to $I_{th}$ using the two temperatures are free parameters yield similar results for all measurements using a given $V_{heat}$. **a** Results from the RTD theory fits to $I_{th}(V_G)$ for all measurements with $V_{heat}$ = 1000 mV and $R$ = 1.5 M$\Omega$. **b** Resulting RTD theory fits to $I_{th}(V_G)$ using $V_{heat}$ = 1000 mV and different $R$ as indicated. **c** The temperatures from fits to all measurements including $V_{heat}$ in a range between 500 and 1250 mV, and $R$ in a range between 14 k$\Omega$ and 250 M$\Omega$.

### III. Validity of the model and heat flow estimate

In order to estimate the heat-to-electric energy conversion efficiency $\eta = P_{th}/J_Q$, our analysis required a quantitative estimate of the heat flow $J_Q$. To achieve this, we used microscopic modeling based on the RTD theory, described in Section C. This theoretical approach includes full nonlinear effects, strong electron-electron interactions and properly takes into account all second order tunneling processes. Furthermore, our QD model has only a few experimentally determined parameters ($\Gamma$ and gate coupling are extracted from $G$, and the $T_{C,H}$ at each $V_{heat}$ from $I_{th}$).

By using the theory described in Section C we found an excellent agreement between the calculated charge current, eq. 21, and the measured $I_{th}$ over wide value ranges of $V_G$, $V_{heat}$ (see fig. 2.c in the main paper) and $R$ (see fig. S10.b). This quantitative and qualitative agreement provides a strong indication that our model fully captures the physics of the experiment. Even though we used modelling – rather than a direct measurement – to estimate the electronic heat flow $J_Q$ using eq. 25 we are confident about the quantitative accuracy of our calculations. An alternative, model-independent way of determining the electronic heat flow, used in Refs. 25 and 26 does not allow for independent control of the thermal and electric biases and is therefore incompatible with our performance analysis at constant $\Delta T$.

## IV. Finding the best fit

When fitting the measured current and conductance to the RTD theory in order to obtain the tunnel rate and the temperatures, the best fit is found by minimizing the sum of squares of the residues

$$r = \sum_i [I_{theory}(V_{G,i}) - I_{experiment}(V_{G,i})]^2. \tag{27}$$

An example of such a residue is shown in Fig. S11 where the current is measured for $V_H$ = 1000 mV and R=14.4 k$\Omega$. The termination condition used for the temperature fits is that the minimum is found with a resolution of 0.003 K and when fitting the tunnel rate the resolution is 0.01 GHz.

Since the residue in Fig. S11 shows one clear, global minimum the fit results in two unique temperatures and the fitting procedure is a valid thermometry tool.

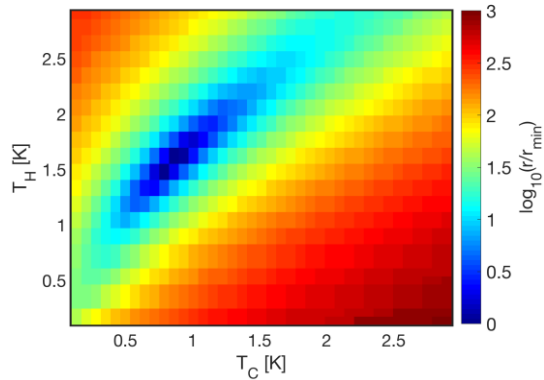

**Supplementary Fig. 11: Residues from temperature fit.** Logarithm of the sum of residues (see eq. 27) normalized by its minimum value when fitting the RTD theory to the measured thermocurrent using $V_{heat}$ = 1000 mV and R = 14.4 k$\Omega$. The best fit is given by $T_C$ = 0.89 K and $T_H$ = 1.46 K. A global minimum means that the fitting technique produces two unique temperatures.

## V. Obtaining the maximal power value from the data

In Fig. 3a,b of the main text $P_{th}$ was calculated using $P_{th} = I_{th}^2 R$ on a point-to-point basis. A polynomial fit was used to obtain the maximal power value from the each $I_{th}$ trace. The gate position of the $I_{th}$ extremum was then used to calculate the electronic heat flow $J_Q$ and thus also the efficiency at maximum power, $\eta_{Pmax}$.

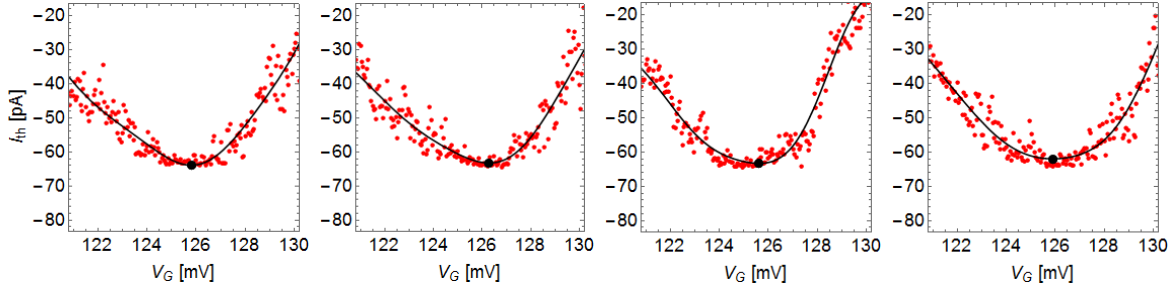

**Supplementary Fig. 12: Obtaining the $I_{th}$ maximal amplitude for the maximal $P_{th} = I_{th}^2 R$.** All four traces show $I_{th}$ data around the maximal power peak for $R = 1$ M$\Omega$ and $V_{heat} = 1250$ mV. In total 400 data points (red dots) around the $I_{th}$ peak are fitted using a polynomial (black line). The peak value and position in $V_G$ is determined by the extremum of the fit (black dot).

### E. Symmetric tunnel rates

The procedure by which we determined $\Gamma$ from conductance data is described in Section D.I. During the fitting process we assumed that the QD is coupled equally strong to both reservoirs, i.e. $\Gamma = \Gamma_C = \Gamma_H$. Here we motivate why we are safe to assume symmetric and energy independent tunnelling rates.

Performing the RTD theory fits to the conductance $G(V_{ext} = 0)$ alone does not provide any information about any asymmetry in tunnelling rates, because the conductance scales with the parameter $\Gamma_C\Gamma_H/(\Gamma_C+\Gamma_H)$ and does not directly depend on $\Gamma_C$ and $\Gamma_H$. Information about asymmetry is however found in the saturation current, i.e. the maximum current the corresponding resonance can carry in either direction. This is easiest seen when gating the device to the resonant condition ($V_G \sim 0.13$ V in Fig 2a of the main text) and applying a large enough bias $V_{ext}$ of both polarities. For example, for a two-fold spin degenerate QD energy level, one finds that a large asymmetry, $\Gamma_{C/H} \gg \Gamma_{H/C}$, results in a factor two for the difference in the saturation currents in opposite directions [27].

In Fig. S13 we compare the measured saturation currents $I(V_{ext})$ in our device to the theoretical predictions using $\Gamma_C = \Gamma_H$, $\Gamma_C = 2\Gamma_H$ and $\Gamma_C = 10\Gamma_H$ while keeping $\Gamma_C\Gamma_H/(\Gamma_C+\Gamma_H)$ fixed. The measured plateau current is not constant, which we attribute to energy dependent density of states (DOS) of the short nanowire leads segments. However, we note that the scale within which these variations manifest themselves is on the order of several mV while the voltages that were developed across the QD during the thermoelectric experiments were limited to much smaller values, less than 0.25 meV, within which it is safe to assume an energy-independent DOS.

The saturation current levels in Fig. S12 suggest that there might be a small $\Gamma$ asymmetry in our device corresponding to $\Gamma_C/\Gamma_H$ between 1 and 2, but it is not possible to obtain a good quantitative estimate of its magnitude. Such an asymmetry will slightly change the thermally induced current and thus also the temperatures extracted form fitting the RTD theory to the current, and more importantly increase the heat current since $\Gamma_H + \Gamma_C$ is increased. In Fig. S14 we show the same type of analyses as in Fig. 3c in the main paper with the exception that the analysis is performed using an asymmetry factor of 2. The difference between $\eta$ at maximum $P$ obtained when using $\Gamma_C/\Gamma_H = 2$, instead of $\Gamma_C/\Gamma_H = 1$, is at most $0.02\eta_C$, showing that the asymmetry has little effect on the analysis of our dot and it is safe to assume symmetric couplings.

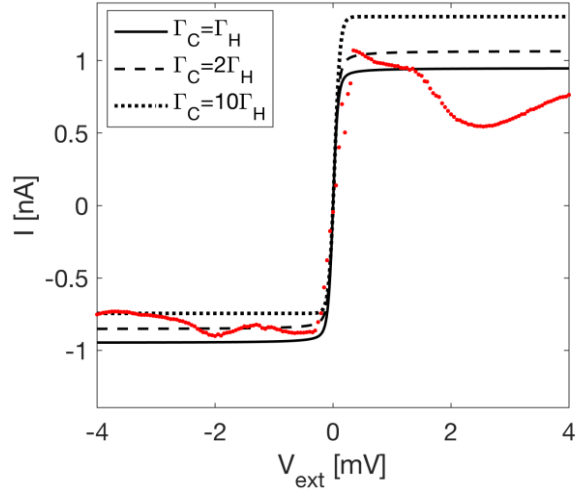

**Supplementary Fig. 13: Saturation current.** Measured (red dots) and calculated (lines) current as a function of $V_{ext}$ at the resonance ($V_G$ = 0.126 V). The RTD calculations use different asymmetries as indicated in the figure. The magnitude of the saturation current is used to probe tunnel coupling asymmetry.

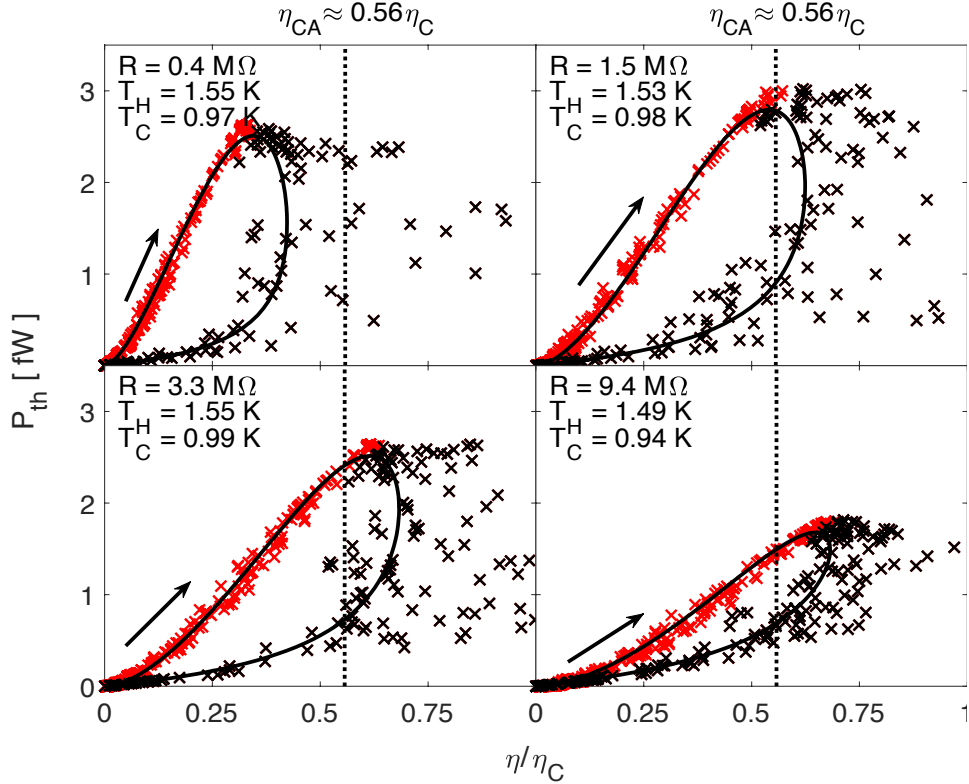

**Supplementary Fig. 14: Thermoelectric performance of the QD including tunnel rate asymmetry.** Same as Fig 3 c in the main paper but using $\Gamma_C = 2\Gamma_H$, $\Gamma_C\Gamma_H/(\Gamma_C+\Gamma_H) = 4.44$ GHz. Parametric plot of $P_{th}$ and $\eta = P_{th}/J_Q$ as $V_G$ is varied at $V_{heat}$= 1000 mV for four different $R$ = 0.4, 1.5, 3.3 and 9.4 M$\Omega$. Red and black markers identify data points from the corresponding $V_G$ ranges as indicated in FIG 3 a,b in the main paper. Data points are based on the measured values of $P_{th}$ and the calculated $J_Q$ using the experimentally determined parameters. The solid lines are based purely on RTD calculations using the same parameters. The arrow indicates the direction for increasing $V_G$. The plots show close resemblance to those in Fig. 3c. The difference between $\eta$ at maximum $P$ obtained when using $\Gamma_C/\Gamma_H = 2$, instead of $\Gamma_C/\Gamma_H = 1$, is at most $0.02\,\eta_C$, which is within the measurement uncertainties.

Any additional resistance in series with the QD would modify the Coulomb peak, which would in turn lead to an incorrect $\Gamma$ value obtained from the $G$ measurements. However, since the magnitude of saturation currents is supposed to be independent of the series load it can be used to check whether any additional resistances, such as contact resistances between the nanowire and the metallic leads or nanowire lead resistances, are negligible. This is done by comparing theory predictions for saturation currents given the obtained $\Gamma$ value. Indeed, from Fig. S12 we see that $\Gamma = 8.9$ GHz, as determined from $G(V_{ext} = 0)$, predicts the saturation current magnitude quite well, and we thus conclude that there are no significant contributions from parasitic resistances.

Further evidence for the negligibility of the contact resistances is provided by a comparison of the measured and calculated thermoelectric power produced by the dot. Our measurement of $P_{th} = I_{th}^2 R$ is determined using the known load $R$. Any additional power dissipated, say in the nanowire leads (a few k$\Omega$ at most), is not accounted for and would lead to an error in $P_{th}$ proportional to the relative contribution of the parasitic resistance to the overall resistance. If this error was significant it would manifest itself as a significant deviation from the predicted power at smaller loads (Fig. 4a). However, this is not the case, further confirming the negligibility of parasitic resistances.